\documentstyle[epsf,12pt]{article}
\def\lsim{\mathrel{\mathpalette\Oversim<}}
\def\gsim{\mathrel{\mathpalette\Oversim>}}
\def\Oversim#1#2{\lower0.5ex\vbox{\baselineskip0pt\lineskip0pt%
            \lineskiplimit0pt\ialign{%
          $\mathsurround0pt #1\hfil##\hfil$\crcr#2\crcr\sim\crcr}}}

\makeatletter
\def\ps@myfrontpage{\let\@mkboth\@gobbletwo
 \def\@oddhead{\large\sl\vbox to0pt{\hbox{\it Department of Physics}
               \hbox{\it Kyoto University}\vss}\hfil
              \rm\vbox to0pt{\pubnumber\hbox{\today}\vss}}%
 \def\@oddfoot{}\def\@evenhead{}%
 \def\@evenfoot{}\def\sectionmark##1{}\def\subsectionmark##1{}}
\def\@maketitle{\newpage
\null
\vskip\temp
 \vskip 2em           
 \begin{center}
  {\LARGE \@title \par}     
  \vskip 3em 
  {\large                        
   \lineskip .5em           
   \begin{tabular}[t]{c}\@author 
   \end{tabular}\par}                   
\end{center}
 \par
 \vskip 3em 
}
\makeatother

\def\begineq{\begin{equation}}
\def\endeq{\end{equation}}
\def\pubnumber{\hbox{KUNS 1294}\hbox{astro-ph/XXXXXXXX}}
\newdimen\temp \temp=0pt
  \setbox1=\vbox{\pubnumber}\advance\temp by 1.5\ht1
\begin{document}
\title{
The fragment mass scale of the primordial gas clouds I.\\
non-spherical  pressure-free collapse
}

\author{Hajime Susa, Hideya Uehara and Ryoichi Nishi\\
{\em Department of Physics, Kyoto University} \\
{\em Kyoto 606-01, Japan}}
\maketitle\thispagestyle{myfrontpage}
\vfill
\begin{abstract}
We investigated the thermal evolution of non-spherical primordial gas clouds 
of mass $ M=10^6M_\odot $. 
We studied two cases: 1) prolate and oblate clouds without angular momentum,
2) initially spherical, oblate and prolate clouds with angular momentum.
In the spherical case, 
the magnitude of the angular momentum is the key quantity
which determines the fragment mass.
The fragment mass is found to be $M \gsim 100 M_\odot$ for 
$10^{-2} \lsim \lambda \lsim 1$, 
where $\lambda$ is the cosmological spin parameter.
For the oblate shape of initial gas clouds, 
the angular momentum is almost always not important, 
In the case of prolate clouds with some angular momentum,
collapse proceeds in the same way as spherical case,
and the bounce occurs to form a disk due to the angular momentum.
In any case, the estimated ``fragment mass'', must be above $100 M_\odot$
for the primordial gas clouds 
with typical angular momentum and typical oblateness or prolateness.
\end{abstract}
\vfill
\newpage
\section{Introduction}
The formation of galaxies or pregalactic objects
in the early universe ($ z \sim 10 - 100 $) is one of the most important
topics. 
Many works have been done on these topics by means of 
semi-analytic and statistical methods 
(Rees \& Ostriker 1977; Efstathiou \& Rees 1988; 
Haehnelt \& Rees 1993; Kaufmann {\it et al.} 1993)
 or huge numerical simulations 
(e.g. Cen \& Ostriker 1995a, 1995b ; Yamashita 1993).
But in these previous works, their "galaxies" are too simple to
explain the very rich features of galaxies, 
especially {\it the existence of stars in the galaxies}.
Some authors have argued about the formation of stars 
within the framework of galaxy formation ({\it e.g.}Katz 1992).
But their model of star formation, 
which assumes that star formation rate is simply proportional to $\rho^n$,
where $\rho$ is the gas density,
and initial mass function is constant, seems to be too simple. 
Star formation processes in the pregalactic cloud may be much different 
from those in the present galaxy 
because of different chemical composition and environment.


We wish to clear the physical processes of the galaxy formation, 
evolutional processes 
of the cosmological density 
perturbations, including star formation.
To make a first step to fix this physics,
we investigate the collapsing process of 
the pregalactic gas clouds with no metals, and discuss the fragment mass.

\par

Many authors have argued about the thermal evolution of the primordial 
gas clouds in pressure free collapse 
(Matsuda {\it et al.} 1969; Hutchines 1976; Carlberg 1981; 
Palla,  Salpeter \& Stahler 1983 (here after PSS)).
As was discussed by Carlberg (1981), collapsing spherical primordial cloud has 
characteristic Jeans mass scales: 
$20 M_\odot$ at ${\rm H}_2$ dissociation, 
and $0.06 M_\odot$ at H ionization. 
But his calculation did not include the three body reactions 
which produce hydrogen molecules, the main coolant of the
primordial gas clouds.
PSS has included the three body reactions, and have shown that 
Jeans mass reaches a temporary minimum and a true minimum values
at ${\rm H}_2$ dissociation and ${\rm H}$ ionization,
and they are smaller than those calculated in Carlberg (1981).
However, these calculations were restricted to 
the case of spherical cloud, 
though spherical cloud in pressure-free collapse
is unstable against non-spherical perturbation (Lin {\it et al.} 1965). 
\par  
As was discussed by Hutchines (1976), 
the thermal evolution of the
clouds are affected so much by the initial shape of the clouds,
or their angular momentum at the initial state.
However, Hutchines (1976) did not include the effect of vibrational
transitions of hydrogen molecule, and the three body reactions.
If $T\gsim 1000 $ K, ${\rm H}_2$ vibrational cooling is comparable
with rotational cooling (e.g. Hollenbach \& McKee 1979).
And if $n \gsim 10^8 {\rm cm}^{-3}$, ${\rm H}_2$ formation rate 
through three body reactions is so high that almost all hydrogens 
become ${\rm H}_2$ (PSS).
So the cooling rate of his calculation was seriously underestimated.
In case of the spherical collapse this difference 
changes the thermal evolution qualitatively (PSS).
\par 
{\it In this paper}, we perform the calculation of non-spherical evolution, 
which include the vibrational transition of ${\rm H}_2$ 
and three body reactions.
We are interested in the destiny of    
the primordial gas clouds, which have no spherical symmetry 
and have some angular momentum induced by the tidal interaction 
between the clouds when they were only density perturbations.
Moreover, in the spherical calculation, we can only derive the Jeans scale, 
but it is not always the scale of fragmentation.
However, in such a non-spherical calculation, 
especially for the collapse of oblate clouds, 
we can estimate the size of the fragments 
from the final disk thickness.  
Then, this calculation has the possibility to estimate the mass of the stars 
which was formed for the first time.
\par
In the next section we will introduce the details of our calculations.  
And in later sections, we will show our results, and discussions.

\section{The collapse model}
In this section, we will show how we treat the dynamics of the collapse, 
chemical reactions, cooling of the cloud due to the hydrogen molecules.

\subsection{Density evolution of the cloud}
 We regard the evolution of the gas cloud 
to be dust collapse. 
This can be justified, if the free-fall time  
$t_{grav}\sim {(G\rho)}^{-1/2}$ is much shorter than the sound crossing time.
And in our calculation, this condition is always kept. 
%
We also assume the cloud to be homogeneous for simplicity.
If we consider
the spherical collapse, 
the evolution of the density is 
easily determined by one equation of motion.
But here, we are interested in the non-spherical collapse of 
the cloud. So we take into account the non-sphericity by using the  
the spheroidal uniform collapse as was introduced by
Lin {\it et al.} 1965 and Hutchines 1976. 
According to Hutchines(1976), a spheroidal 
collapse with angular momentum, can be described as,\\
  
\begin{eqnarray}
{{d^2 R}\over{d {\tau}^2}}& = 
 & -\alpha_0 F(\alpha) R^{-2} + \xi R^{-3},\label{eqn:drdt}\\
{{d^2 Z}\over{d {\tau}^2}}& =
& -{\alpha_0}^{-1} G(\alpha) R^{-1}Z^{-1}, \label{eqn:dzdt}\\
 \rho & =  & \rho_0 R^{-2} Z^{-1}, 
\end{eqnarray}
where $R,Z$  are the radial and axial size of the cloud normalized 
by the initial size. And also other variables or functions are 
defined as follows.\\
\begin{eqnarray}
r & \equiv & r_0 R, \\
z & \equiv & z_0 Z, \\
\alpha_0 & \equiv & z_0/r_0, \\
\alpha & \equiv & z/r = \alpha_0 Z/R, \\
\tau & \equiv & t/{(4\pi G \rho_0)}^{-1/2},\\ 
\xi & \equiv & {(\omega_0 {(4\pi G \rho_0)}^{-1/2})}^2,
\end{eqnarray}

where, $r_0,\  z_0$ are initial physical size of the cloud,
$\rho_0$ is the initial density, $t$ denotes the time measured from 
the beginning of the collapse, and $\omega_0$ is the initial angular 
velocity of the  cloud. Definitions of two functions $F(x), 
G(x)$ are,

\begin{eqnarray} 
F(x) & \equiv & 
{{\arccos(x) - x \sqrt{(1-x^2)}}\over{2(1-x^2)^{3/2}}},\\
G(x) & \equiv & 
{{-x^2 \arccos(x)+x \sqrt{(1-x^2)}}\over{(1-x^2)^{3/2}}},
\end{eqnarray}
for $x < 1$, and
\begin{eqnarray} 
F(x) & \equiv & 
{{\ln(x+\sqrt{x^2-1}) - x \sqrt{x^2-1}}\over{2(x^2-1)^{3/2}}},\\
G(x) & \equiv & 
{{-x^2 \ln(x+\sqrt{x^2-1})+x \sqrt{x^2-1}}\over{(x^2-1)^{3/2}}},
\end{eqnarray}
for $x > 1$.

Equations ($\ref{eqn:drdt}$) and ($\ref{eqn:dzdt}$) are solved numerically, 
with the initial conditions,
\begin{eqnarray}
R(0) & =&  Z(0) =1,\\
{{d R}\over{d {\tau}}}&  = &{{d Z}\over{d {\tau}}}\ \  = 0.
\end{eqnarray}
These initial conditions correspond to that of 
the density perturbations 
at their maximum expansion epoch.

\subsection{Thermal evolution and chemical reactions}
Here we will show the energy equation and the cooling functions.
We also summarize the reaction rates for five compositions which 
we take into account for this calculation.
\par
First of all, the following is the energy equation which should be solved:
\begin{equation}
{{d \epsilon}\over{dt}} = p(T,\rho) {{d\left( 1/\rho \right)}\over{dt}}
- \Lambda(\rho,T),
\end{equation}
where $\epsilon$, $p$, $\rho$ and $\Lambda$ denotes 
the internal energy per unit mass, 
the pressure of the gas cloud, the density of the cloud and 
the emitted energy per unit mass per second from the gas cloud, respectively. 
We also use the equation of state of ideal gas.
The first term on the right hand side of this equation 
denotes the adiabatic heating, 
and the second term, $\Lambda$, 
denotes the cooling rate or heating rate 
due to various physical processes excluding the adiabatic heating.
\par
In $\Lambda$ we take into account the ${\rm H_2}$ vibrational and rotational 
line cooling, and the heating and the cooling due to the ${\rm H_2}$
formation and dissociation (Appendix A).
We also calculated the radiative transfer equation to calculate the 
radiative part of $\Lambda$. 
To know the radiated energy from the cloud, 
we calculate the net flux radiated from the surface of the cloud 
(Appendix B).
Then, we obtain the radiative cooling rate per unit mass of the cloud.
\par
The reaction rates are taken from the Table 1. of  PSS.
The chemical compositions 
which we introduced to this calculation are 
${\rm e}^-, {\rm H}^+, {\rm H}, {\rm H}^-, {\rm H}_2$.
We exclude helium for simplicity,
since helium play an important role only at high temperature 
$(T > 10^4 {\rm K})$.

\subsection{Initial conditions}
Now we can calculate the thermal evolution of the collapsing 
gas clouds with a given set of initial conditions. 
In this paper, we set the mass of the cloud to be $10^6 M_\odot$,
which is consistent with the Jeans mass just after the recombination.
We set other physical quantities as 
$\rho_0 = 4.6\times 10^{-24} {\rm g cm^{-3}}$,
$T_0=10^2 {\rm K}$, $f_e=f_{\rm H^+}=2.8\times 10^{-5}$, $f_{\rm H^-}=
3.0\times 10^{-12}$, $f_{\rm H_2}=6.3 \times 10^{-6}$, respectively.
Here $\rho_0$, $T_0$, $f_X$ denotes the initial density, 
the initial temperature, 
and the number fraction of the composition $X$, respectively.
For comparison, we choose the same values as PSS's, 
and they are also close to those used by Hutchines (1976).
Then, we have two free parameters for this calculation, 
$\alpha_0$ and  $\xi$. $\alpha_0$ denotes the flatness of the initial spheroid,
and $\xi$ denotes the ratio of the energy of the rotation of the cloud 
to the gravitational energy of the cloud.
We will show the results for different values
of these two parameters, in next the section.

\section{Thermal history of the collapse}

\subsection{The collapse of the non-rotating clouds : $\alpha_0\ne 0, \xi =0$}
The collapse of non-rotating spheroid can be  divided into two types,
according to their initial conditions.
\par
First, the collapse of a initially oblate cloud is very different 
from that of spherical one, as was described by Hutchines (1976).
The non-spherical collapse at the initial phase is immediately 
followed by the disk like collapse, which has a different time scale
from a spherical collapse. 
Strictly speaking, the time needed for $\rho$ being $\infty$, 
are not much different from each other,
but {\it at the same density}, which is much higher than the initial value, 
the time scales simply determined 
by the temporal shrinking rates are entirely different 
each other, as
\begin{eqnarray}
t_{dyn} \equiv -r/v_r = 
\left(3\rho/{{d\rho}\over{dt}}\right) &  \propto & \rho^{-1/2}
({\rm sphere}), \label{eqn:sphere}\\
\equiv \; -z/v_z =\;  \left(\rho/{{d\rho}\over{dt}}\right) 
&  \propto & \rho^{-1} ({\rm sheet}).\label{eqn:sheet} 
\end{eqnarray}

Because of the thinness of the disk,
$t_{dyn}$ of sheet like collapse is much smaller 
than the free-fall time at the same density.
And this time scale, $t_{dyn}$ is also the time scale of adiabatic heating.
So the thermal evolution of these oblate spheroids are also 
very different from that of spherical clouds.
The difference is shown in Fig.1, 
which is essentially the same as Hutchines (1976). 
Then we can conclude that in such a type of collapse,
the three body reactions are not important for the thermal history 
of the cloud. In other words, the temperature of the cloud rises
so fast that the density can not rise to the critical density 
($\sim 10^8 {\rm cm}^{-3}$), which is sufficient for the three body 
reactions to become a main  contributor of producing hydrogen molecules.
\footnote{
Of course, sufficiently small $\alpha_0$ provides the opposite results,
but the value $\alpha_0$ need to be extremely close to unity, 
say 0.9999.
}
\par
Second, the collapse of a initially prolate cloud (Fig.2) is not so different
from that of the spherical one,
because they have almost the same collapsing time scale, 
%
\begin{eqnarray}
t_{dyn} &  \propto & \rho^{-1/2} {(\ln{\rho})}^{-1/2} 
({\rm cylinder}),\label{eqn:cylinder}
\end{eqnarray}
as the spherical one.
Then we can see the time scale $t_{dyn}$ is not 
so different from that of the spherical one, 
so the thermal evolution of the non-rotating prolate cloud
is very similar to that of the spherical cloud.
The only difference is the density to be opaque 
to the line emission by ${\rm H_2}$..
In the case of prolate collapse, 
is larger than the spherical case, because of its shape.

So we can see that in this case, the three body reactions can be
important. 
However, the cloud shape turns out to be a cylinder, different from
the spherical case. 


\subsection{The collapse of rotating clouds}
\subsubsection{\it Initially spherical rotator : $\alpha_0 =0, \xi\ne 0$}
We also interested in the effects of angular momentum. 
To pull out its effects independently from that of the initial shape of the cloud,
we test the model of initially spherical rotator (Fig.3).
The initially spherical rotator experiences two phases during their free-fall 
collapse.
\par
First, the cloud collapses spherically, because of its initial configuration.
But when the size of the cloud shrinks enough for the angular momentum barrier
to stop the radial infall of the gas, the shape becomes disk like.
 Then in this second phase, eqs. ($\ref{eqn:sphere}$),($\ref{eqn:sheet}$) 
tells that 
the dynamical time scale is shorter than in the spherical case 
at a same density.
And what is more, this time scale 
is much shorter than the cooling time scale in our calculation. 
So the epoch at which the type of the collapse switches 
from spherical type to disk like one
has a significant importance on the later thermal evolution of the cloud. 
For example, the three body reactions,
\begin{eqnarray}
 3{\rm H}&\rightarrow & {\rm H}_2+{\rm H},\\
 2{\rm H}&+&{\rm H}_2 \rightarrow 2{\rm H}_2, 
\end{eqnarray}
have larger reaction rates for larger density. 
These reaction rates exceeds the two body reaction rates for $n\gsim 10^8 
{\rm cm^{-3}}$. 
But if the cloud shrinks disk like and the temperature
increase almost adiabatically during $n\lsim 10^8{\rm cm^{-3}} $, 
the collapse will be stopped by the thermal pressure 
before the three body reactions become effective.
This situation is realized for large angular momentum.
On the contrary, if the collapse proceeds spherically during 
$n\lsim 10^8 {\rm cm^{-3}} $,
the three body reactions could play important roles before the the cloud
 is stopped by the thermal pressure.

We can find the critical value of rotation parameter $\xi$, 
which is sufficiently so small that the three body reactions become 
important for the formation of hydrogen molecules.

This critical value of $\xi$ can be roughly obtained by the following estimate.
According to the eqs. ($\ref{eqn:drdt}$),($\ref{eqn:dzdt}$), 
$ R_{sw}$, 
the radius when the collapse type switches from spherical collapse to disk like
 one is simply determined as $R_{sw} = \xi$. 
Then, using the mass conservation law for the spherical collapse, 
the corresponding density is determined as 
$ n_{sw} R_{sw}^3 = n_{sw}\xi^3 =  n_0 $.
Finally, we can obtain the critical value of $\xi$ 
from the equation $n_{sw} \simeq n_{3b}$, 
where $n_{3b}$ is the critical density 
above which the three body reaction rates exceed the two body reaction rates.  
Substituting $n_0$ for $1{\rm cm^{-3}}$ and  
$n_{3b}$ for $10^{-8}{\rm cm^{-3}}$, 
We obtain $\xi \sim 10^{-3} $.
Actually in Fig.4, the maximum fraction of ${\rm H_2}$ 
for different value of $\xi$  
is suddenly changed at $\xi \sim 10^{-4} - 10^{-3}$.
But the hydrogen molecules, produced through these three body reactions,
cannot radiate the thermal energy away from the cloud freely 
because of the absorption by itself.
In other words, the rapid increase of the hydrogen molecules causes 
the rapid increase of 
the optical depth of the line emission of ${\rm H_2}$. 
As a consequence, the effect of three body reactions are not very important
as for the energetics of the gas cloud.
Of course, this situation is different if the mass scale of the 
cloud is much smaller than $10^6 M_\odot$.

We can also mention that this value is very close to 
the cosmological value of $\xi$ ($\simeq 10^{-4}\sim 10^{-3}$),
originated from the tidal interactions among the density perturbations during 
their linear phase.
Many authors have discussed that the orders of magnitude of cosmological 
spin parameter $\lambda$ almost equals to 
$0.03$ (e.g. Eisenstein \& Loeb 1995 ).

Of course this value is obtained for much larger mass scale 
than $10^6 M_{\odot}$, which we are interested in now, 
but many results support that
$\lambda$ is almost independent on the mass scale 
(Eisenstein \& Loeb 1995 see also references there in) .
And we can easily check the relation between $\lambda$ and $\xi$ is 
 $\lambda \simeq \sqrt{\xi}$.     
This means the possibility that the angular momentum of the cloud 
can be a essential parameter which determines the fragment mass.
\subsubsection{\it Initially non-spherical rotator : $\alpha_0\ne0, \xi\ne 0$}
Here we will show the thermal evolution of initially non-spherical rotator.
This results can be explained as a combined results
of the above two sections. 
In the case of initially oblate rotator,
with some realistic values of initial oblateness ($\alpha_0 \sim 0.9$) (Bardeen et.al. 1985) 
and of angular momentum ($\xi \sim 10^{-3}$),
the cloud completes its collapse before the angular momentum barrier prevent
the gas from radially collapsing. 
So in this case, the essential parameter is not $\xi$, but $\alpha_0$.
\par
In contrast, in the case of initially prolate rotator (Fig.6), 
$\xi$ is an essential parameter,
because of the faster collapse in radial direction than the z-direction.
The gas falling in radial direction 
inevitably encounters the angular momentum barrier, and then bounces.
The bounced cloud is adiabatically cooled in the beginning, 
then recollapses to be adiabatically heated up.
But when the cloud recollapsed to the density at which it bounced before,
the shape of the cloud is no more prolate, but oblate.
And the cloud collapses to a disk, immediately 
followed by the end of the calculation. 
\par

\section{The mass of the fragments}

As mentioned in the former sections, 
we can calculate the basic equations with any initial conditions.
But the dynamical equation which we introduced to solve this problem, 
does not contain the thermal pressure force.
In other words, if the condition
\begin{eqnarray}
(4 \pi G\rho)^{-1/2} & =  min(r,z)/c_s \label{eqn:fin},
\end{eqnarray}
is satisfied, our dynamical equation breaks down.
Then, at this time we stop our calculation. 
and we estimate the mass of the fragments as follows. 
For example, in the case of initially spherical rotator, 
the evolution of the Jeans mass and 
$M_{mgr}(t)\equiv \rho \cdot 2z \cdot \lambda_{mgr}^2 $
is shown in Fig.5, where $\lambda_{mgr}$ is the wave length for which 
the perturbation grows up fastest. 
We estimate $\lambda_{mgr}\simeq 2 \pi \cdot 2z$ (Larson 1985) 
in this paper.
In Fig.5, the gravitational force and the pressure gradient force balances
where the two lines crosses each other, and our calculation is stopped 
at this point. 
We estimate the mass of the fragments with the definition, 
$M_{frag} \equiv M_{mgr}(t_f)$, 
where $t_f$ is the time that the condition ($\ref{eqn:fin}$) is satisfied.   
 We also calculated the mass of the fragments 
for wide variety of initial conditions (Fig.7). 
It is obvious that the spherical symmetry is a very special configuration
as for the thermal evolution of a collapsing cloud.
Three body reactions to produce hydrogen molecules 
are only effective when the initial shape is extremely spherical or 
the initial configuration is prolate, and has extremely low angular momentum. 
\par
In the region that the cloud is oblate ($\alpha_0 < 1$),
the fragment mass has almost no dependence on $\xi$ for small $\xi$. 
This can be understood naturally as follows: 
The deformation effect simply by the gravitational instability, 
exceeds the effect by the angular momentum barrier for small $\xi$.
In other words, the disk collapses to a very thin disk before the bounce
due to the angular momentum occurs.
\par
In case the cloud is prolate ($\alpha_0 > 1$),
the clouds whose angular momentum vector 
is parallel to the major axis of the cloud,
finally form a thin disk whose radial gravity 
is supported by the angular momentum barrier, as was mentioned in section 3.2.
We can see that 
there is a ``trough''on $\alpha_0 - \xi$ plane 
at $10^{-1.5}\lsim \xi \lsim 10^{-1}$,
 which corresponds to the value $\xi_E, \xi_B$ 
described by the equation (41) in Hutchines (1976). 
If $\xi > \xi_B(\alpha_0)$ , 
no bounce occurs by the angular momentum barrier, 
and if $ \xi > \xi_E(\alpha_0)$, 
the density does not decrease after the bounce.  
The ``trough'' structure is understood as follows:
if $\xi > \xi_E(\alpha_0)$, the collapse is almost the same 
as the disk collapse, so $M_{frag}$ is larger than 
that in the case  $\xi < \xi_E(\alpha_0)$.
If $\xi_E(\alpha_0) > \xi > \xi_B(\alpha_0)$, 
the density always increases, 
, the surface density is also 
increases, and $M_{frag}$ decreases.
It is different from that in case the disk collapse.
Finally, if $\xi < \xi_B(\alpha_0)$, 
due to the strong bounce in $R$ direction,
the density decreases, 
and  $M_{frag}$ is larger than that in the previous case.
But if $\xi$ is small enough, ${\rm H_2} $ is produced during the bounce,
and the corresponding cooling proceeds. 
Then $M_{frag}$ is smaller than that for  $\xi \simeq \xi_B(\alpha_0)$.  
%
\section{Conclusion \& discussion}
We have shown the thermal evolution of 
the collapse of primordial gas clouds with non-spherical initial 
configuration and/or with some angular momentum. 
We found the fragment mass is 
very sensitive to the initial configurations.
\par
In the case of initially oblate clouds, they immediately collapse 
to form a thin disk,
and the gravitational force balances with the pressure gradient force 
before ${\rm H}_2$ dissociation. 
So two characteristic Jeans mass scales at ${\rm H}_2$ dissociation 
and ${\rm H}$ ionization mentioned by Carlberg (1981) 
do not play important roles.  
\par
In the case of initially extremely spherically symmetric clouds,
the evolution of these clouds are affected so much by 
the value of the angular momentum.
If the spin parameter $\xi$ is much larger than some critical value $10^{-4}$
, the dissociation of the hydrogen molecules sets in before the three body 
reactions, which produce hydrogen molecules, becomes effective.
If the spin parameter $\xi$ is smaller than $10^{-4}$, 
the three body reactions can be effective 
before the temperature rises so high that the hydrogen molecules dissociate.
\par
In the case of initially prolate clouds, the angular momentum also plays 
an important role, but the three body reactions are not important for the 
realistic value of $\xi$, or for larger mass scale of the cloud. 
We also calculated the fragment mass $M_{frag}$, 
for various $\alpha_0$ and $\xi$. 
Then the mass is much above the ordinary stellar mass for the realistic 
value of $\alpha_0$ or $\xi$. So we must conclude the initially spherical
configuration is a very special and unrealistic one.
\par
Our assumed configurations, 
in which the angular momentum vector is parallel to 
the axis of the symmetry of the cloud, 
are not realistic, 
because the tidal interaction between the density perturbations 
cannot induce the angular momentum parallel to the axis of symmetry.
And in case we treat the triaxial ellipsoids, 
it is easily understood that 
the angular momentum is not parallel to any axis in general.
In such cases, the angular momentum could play an important role,
because the disk like collapse is stopped by the angular momentum 
which is not parallel to the axis of symmetry.
\par
In this paper, as mentioned in section 4, 
$M_{frag}$ is defined as the Jeans mass at the time essentially 
($\ref{eqn:fin}$)
is satisfied, i.e. we stopped our calculation when the gravitational force
balances with the pressure gradient force for the first time.
But in reality, 
the cloud does not stop dynamical evolution, 
and does not fragment at this time,
since $t_{dyn}$ is much smaller than $t_{ff}$ 
(see equation ($\ref{eqn:sheet}$)).
After the cloud dissipate the kinetic energy, it settle into 
an equilibrium state for z-direction. 
According to Inutsuka \& Miyama (1992), when the condition
\begin{eqnarray}
 t_{dyn} & \equiv & \rho/\frac{d\rho}{dt} \gsim t_{mgr} \label{eqn:crit}, 
\end{eqnarray}
where,\\
\begin{eqnarray}
 t_{mgr}& \simeq &{(\frac{G\rho}{\Omega\pi})}^{-1/2}
\end{eqnarray} 
is satisfied, and if the cooling is effective, 
the cloud will fragment into smaller filaments.
Here, $t_{mgr}$ is the maximum growing rate of the density perturbation 
in the equilibrium sheet, 
where $\Omega$ is a non-dimensional constant ($\sim O(1)$), 
which almost equals 5 in case the sheet is isothermal (Larson 1985).
In this phase, we can argue the true mass of the fragments.
Since the cloud is highly flattened ($\alpha \ll 1$) in this phase,
the radius shrinks little before vertical equilibrium is satisfied,
and the surface density changes little, too.
Thus, the true fragment mass scale is determined by the temperature, 
$T_{fr}$, at the time $t_{fr}$, when equation ($\ref{eqn:crit}$) 
is satisfied.  
In the course of dissipating the kinetic energy,   
the cloud would be shock heated, 
because the collapse of the cloud is supersonic,
and keep cooling during the entire time. 
$T_{fr}$ and the fragment mass scales are increased by the shock heating, 
and decreased by the cooling. 
So it is necessary to investigate the process from $t_f$ to $t_{fr}$,
if we estimate the accurate fragment mass scales.
However, they may not be much different from $M_{frag}$.
Since the shock velocity, $v_s$ is not too high ($v_s \lsim 10 {\rm km/s}$)
 with the parameter of this paper, $T_{fr}$ must be lower than 
$10^{4} {\rm K}$, and  $T_{fr}\gsim 10^3 {\rm K}$
 because the cooling rate of ${\rm H_2}$ is very low for $T < 10^3 {\rm K}$. 
If $v_s \gsim 50 {\rm km/s}$, 
the shock heating would cause a more complicated situation, 
with the ionization of hydrogen atom (Shapiro \& Kang 1987).
In the course of dynamical cooling process, 
the over abundance of electrons is caused 
due to the slow recombination rate compared to the atomic cooling rate. 
And what is more, 
those electrons can be the catalyst for the ${\rm H_2}$ production.
So, serious amount of molecular cooling is 
expected below $10^4{\rm K}$ 
in this dynamical situation (Shapiro \& Kang 1987). 
Then, qualitatively different results may be expected. 
But if we wish to treat this problem properly, 
we will need a one dimensional  hydro dynamical calculation. 
\par
In the second place, the cloud would fragment into filaments 
rather than the blobs according to the nature of gravity. 
Then, the filaments would collapse and bounce to form 
an equilibrium states. 
In this phase, the fragmentation will set in again.
Further discussion will be coming up in our next work.

\par
\vskip 0.5cm
\centerline{ACKNOWLEDGEMENTS}
\vskip 0.3cm    
 We would like to thank H.Sato, T.Nakamura, Y.Yamada and M.Shibata 
for useful discussions. 
We also give our thank to M.J.Rees for his important comments.
This work was supported in part by
the Monbusho Grant-in-Aid for Scientific Research Fund, No.3077.
\pagebreak
\appendix
\centerline{\Large \bf APPENDIX}
\section{The Cooling Processes}
We include various cooling processes to calculate 
the cooling function $\Lambda$.
\begin{description}

\item{1.}Line cooling due to hydrogen molecules \\
Hydrogen molecules have many excited levels 
corresponding to the vibrational and rotational
quantum states.
The radiative transitions 
from a higher energy level to a lower energy level 
of hydrogen molecules 
cools the cloud.
 Then, we have to estimate the level populations of $\rm{H_2}$,
so that we can calculate the emissivity 
$j_\nu \;{\rm[\;erg\cdot s^{-1} \cdot cm^{-3} Hz^{-1}]}$.
 The level populations are determined 
by the statistical equilibrium of the transitions between those states.
To know the level populations, we need the Einstein's $A$ coefficients
between the levels, and also need the collisional excitation rates.
We take those values from  
Hollenbach et.al (1979) and Turner et.al.(1977), respectively. 
Once we know the level populations, 
we can write the emissivity of a energy transition as follows:
\begin{equation}
j_\nu=\Delta E_{i \rightarrow j} A_{ij} n_i,
\end{equation} 
where $\Delta E_{i \rightarrow j},A_{ij},n_i$ denote 
the energy difference of the two states, Einstein's $A$ coefficient,
and the number density of the $i$th state of ${\rm H_2}$.

Using this emissivity and the level populations, 
we can solve the radiative transfer equation for a line emission. 
The derivation of the cooling function is shown in the Appendix B.


\item{2.}${\rm H}_2$ dissociation \\

Hydrogen molecules have lower potential energy than the state 
which it is separated into two neutral hydrogens. 
So the dissociation of hydrogen molecule absorbs the thermal energy
of the another particle of which impact against hydrogen molecules
cause the dissociation.

The cooling rate is written as follows :

\begin{equation}
 \Lambda_{diss} = -7.16\times 10^{-12} \left(\frac{d n({\rm H}_2)}{dt}\right)_- 
{\rm [\;erg \cdot s^{-1}\cdot cm^{-3}]},
\end{equation} 

where $n({\rm H}_2)$ denotes the number density of hydrogen molecules.
In the above equation, $(d n({\rm H}_2)/dt)_- $ is the dissociation
rate of ${\rm H_2}$ (Shapiro et.al 1987).
The time variation of the fraction of any species at each time step 
is known in our calculation, 
because we also solve the time dependent rate equations
coupled with the energy equation.
\item{3.}${\rm H}_2$ formation \\
This process is essentially the opposite process to 
the ${\rm H}_2$ dissociation process. 
But when the hydrogen molecule form, it is in a excited state.
If the collisional de-excitation process 
is dominant compared to the spontaneous decay process,
the excitation energy flow into the thermal energy of electrons.
But if the spontaneous decay rate 
is greater than the collisional de-excitation rate,
the energy will be radiated away (${\rm H_2}$ line emission),
and this process does not help to heat the gas cloud.
So we need this correction term which should be multiplied
to $\Lambda_{diss}$.
According to Shapiro et.al. (1987),  the heating rate is expressed as,   
\begin{equation}
 \Gamma_{form}= 
7.16\times 10^{-12} \left(\frac{d n({\rm H}_2)}{dt}\right)_+ 
\frac{1}{1+n_{cr}/n_H } 
{\rm [\; erg \cdot s^{-1} \cdot cm^{-3}]},
\end{equation}
where $n_H$ denotes the total number density of hydrogen nuclei,
$(d n({\rm H}_2)/dt)_+ $ is the formation rate of ${\rm H_2}$,
and $n_{cr}$ is the critical density which is defined as follows:
\begin{equation}
n_{cr}=\frac{10^6}{\sqrt{T} n_H}\left( 
1.6 \exp\left({-\left(\frac{500}{T}\right)^2}\right)n({\rm H}) 
+1.4 \exp\left({-\frac{12000}{T+1200}}\right)n({\rm H_2}) \right).
\end{equation}
In the above equation, 
$\left( 1+n_{cr}/n_H \right)^{-1}$ is the correction term,
due to the spontaneous decay of excited levels.
\end{description}


\section{Radiative Transfer}
Here we will show the solution of radiative transfer equation
in the case of uniform sphere or uniform disk.
 
And the derivation of cooling function 
due to the line cooling is shown. 
First, the radiative transfer equation along the $s$-direction 
which we must solve is as follows:

\begin{equation}
 \frac{dI_{\nu} }{ds} = -k_{\nu}I_{\nu}+j_{\nu},            
\end{equation} 

where $I_{\nu},k_{\nu},j_{\nu}$ 
denotes the intensity of the radiation, the opacity, 
and the volume emissivity, respectively.

And this equation is easily integrated,

\begin{equation}
I_{\nu}(\tau (\nu))=I_{\nu}(0)e^{-\tau (\nu)}+
\int_0^{\tau(\nu)}e^{-(\tau(\nu)-\tau')}S_{\nu}(\tau') d\tau',
\end{equation}
where,
\begin{equation}
S_{\nu} \equiv j_{\nu}/k_{\nu},  \;\;\; \tau(\nu) \equiv \int_0^s k_{\nu} ds.
\end{equation} 

If the cloud is uniform, $k_{\nu},j_{\nu},S_{\nu}$ must be constant
along any line of sight.  Then, the solution is simplified as follows.
\begin{eqnarray}
\tau (\nu)& =& k_{\nu} s      \\
I_{\nu}\left(\tau (\nu)\right) &=&S_{\nu}+e^{-\tau (\nu) }\left(I_{\nu}(0)-S_{\nu}\right)
\end{eqnarray}

Now we can calculate the flux radiated from a uniform sphere.
The energy flux is written in terms of the intensity,
\begin{equation}
F_{\nu} = \int d \Omega I_{\nu}(\theta,\varphi) \cos \theta .
\end{equation}
In case we treat uniform sphere, 
the above integration can be performed easily 
with the boundary condition $ I_{\nu}(0)=0$,
and the final result is as follows:

\begin{equation}
F_{\nu} = \pi S_{\nu}
\left(1+
\frac {e^{-2\tau_{tot}(\nu)}\left(1+2\tau_{tot}(\nu)\right)-1}
{2 {\tau_{tot} (\nu)}^2} \right), 
\end{equation}
 where, $\tau_{tot}(\nu)= k_{\nu} R$, and R denotes the radius 
of the uniform sphere that we regard as the cloud.

Then, we can estimate the cooling function 
$\Lambda_{rad}(\nu) 
{\rm [\; erg\cdot s^{-1}\cdot cm^{-3} \cdot Hz^{-1}]}$
 to be,

\begin{equation}
\Lambda_{rad}(\nu) =  \frac{3 F_{\nu}}{R}.
\end{equation}

\par

This treatment is different from PSS, 
in which they regard the cooling rate more approximately as follows:
 for $\tau < 1$, they calculate the cooling rate 
by assuming that all radiation 
emmited from the gas can escape to the outside of cloud, 
and for $\tau > 1$, 
they estimate the flux radiated from the surface of the cloud 
at Planck distribution times Doppler width.  

The difference between our treatment and that of PSS is 
significant at $\tau \simeq 1$. 
And it must be noted that significant amount of total emitted energy 
is radiated away from the cloud around $\tau \simeq 1$. 

We also can integrate the radiative transfer equation 
for a plane parallelly symmetric cloud.
In this case, $F_{\nu}$ can be written analytically 
using the 1st exponential function,

\begin{equation}
F_{\nu} = \pi S_{\nu} \left(\left(1-e^{-\tau_{tot}(\nu)}\right)\left( 1-\tau_{tot}(\nu) \right )
+{\tau_{tot}(\nu)}^2 E_i\left(-\tau_{tot}(\nu)\right)\right),
\end{equation}
where $E_i(x)$ is the 1st exponential integration,
\begin{equation}
E_i(x) \equiv - \int^\infty_{-x} \frac{e^{-t}}{t} dt \;\;\;\; (for \; x<0).
\end{equation}
And $\tau_{tot}(\nu)$ in the above equation is redefined as,
\begin{equation}
\tau_{tot}(\nu) \equiv  k_{\nu} Z,
\end{equation}
and Z denotes the thickness of the disk.

Then we approximate $F_{\nu}$ as the flux radiated 
from the thin disk. 
This approximation is valid if the disk is thin enough.
Then the cooling function $\Lambda_{rad}(\nu)$ is calculated 
as follows.

\begin{equation}
\Lambda_{rad}(\nu) =  \frac{2 F_{\nu}}{Z}.
\end{equation}  

If the cloud can be approximated as a infinitely long cylinder,
 we can calculate  $\Lambda_{rad}(\nu)$ ,
but it cannot be expressed analytically.

In an empirical expression,
\begin{eqnarray}
F_\nu & = & S_{\nu}\left(\pi-\int_0^{2\pi}d\varphi 
\int_0^{\pi/2}d\theta \sin\theta\cos\theta
\exp\left(-{\tau_{tot}(\nu)\frac{\cos\theta}
{{\cos}^2 \theta+{\sin}^2 \theta {\sin}^2 \varphi }}\right) \right) \nonumber\\
      & = & S_{\nu}f\left(\tau_{tot}(\nu)\right), \\
\tau_{tot}(\nu) & \equiv & 2 k_{\nu} R ,
\end{eqnarray}
where
\begin{equation}
\begin{array}{rclr}
\vspace{3mm}

f(x) & = & 2 \displaystyle \pi (x-\frac{2}{3}x^2) & 
({\rm for}\;\; x < 0.01), \\
\vspace{1.5mm}
& = & {\rm dex}\; [\; 0.40793+0.308611y-0.344113y^2 \\
& \; & \hspace{1.5cm} +0.0673944y^3+0.057328y^4 \;] 
        \hspace{-5cm}& ({\rm for} \;\;y=\log_{10}x \;\; 0.01\le x \le 10),  \\ 
\vspace{3mm}
     & = & \pi & ({\rm for}\;\; 10 < x),\\
\vspace{3mm}

\end{array}
\end{equation}
and $R$ denotes the radius of the cylinder.
These expressions are obtained by fitting  
the numerically integrated solution.
In this case, the cooling function is expressed as,
\begin{eqnarray}
\Lambda_{rad}(\nu) & = & \frac{2 F_\nu}{R}. 
\end{eqnarray}
\par
Once we know the expression of $\Lambda_{rad}(\nu)$, 
we can also write the cooling rate 
integrated over the frequency.
The cooling rate due to the line emission of hydrogen molecules,
$\Lambda_{rad} 
{\rm [\; erg\cdot s^{-1}\cdot cm^{-3} ]}$
is expressed as follows:

\begin{equation}
\Lambda_{rad}(\nu) =  \sum_{i,j} \Lambda_{rad}(\nu) \Delta \nu ,
\end{equation}

where $\Delta \nu$ denotes the line width 
determined by the Doppler broadnig.  

\vskip 0.7cm
\centerline{\Large \bf REFERENCES}
\vskip 0.5cm
\begin{description}

\item{}Carlberg,R.G., {\it Mon.Not.R.astro.Soc.}, {\bf 197}(1981), 1021.

\item{}Cen,R., Ostriker,J.P., {\it Astrophys.J.}, {\bf 417} (1993a), 404.

\item{}Cen,R., Ostriker,J.P., {\it Astrophys.J.}, {\bf 417} (1993b), 415.


\item{}Efstathiou, G. \& Rees, M.J., {\it Mon.Not.R.astro.Soc.} , {\bf 230} (1988), 5.
 
\item{}Eisenstein, D.J., and Loeb, A., {\it Astrophys.J.},{\bf 439} (1955), 520. 

\item{}Haehnelt, M.G., and Rees, M.J., {\it Mon.Not.R.astro.Soc.}, {\bf 265} (1993), 727.

\item{}Hoyle,F., {\it Problems of Cosmical Aerodynamics} (International Union of Theoretical and Applied Mechanics, and International Astronomical Union) (1949), p.195.

\item{}Hollenbach,D. \& McKee,C., {\it Astrophys.J. Suppl.}, {\bf 41} (1979), 55.

\item{}Hutchines,J.B., {\it Astrophys.J.}, {\bf 205} (1976), 103.

\item{}Inutsuka,S. \& Miyama,S.M., {\it Astrophys.J.}, {\bf 388} (1992), 392.

\item{}Katz, N. \& Gunn, J.E., {\it Astrophys.J.}, {\bf 377} (1991), 365.

\item{}Larson, R.B., {\it Mon.Not.R.astro.Soc.}, {\bf 214} (1985), 379.

\item{}Lin, C.C., Mestel, L., and Shu, F.H., {\it Astrophys.J.}, {\bf 142} (1965), 1431.  

\item{}Takeda, H., Sato, H. \&  Matsuda, T., {\it Prog.Theor.Phys.}, {\bf 42} (1965), 219.

\item{}Turner, J., Kirby-Docken, K. \& Dalgarno, A., {\it Astrophys.J. Suppl.} , {\bf 35} (1977), 281.

\item{}Palla, F., Salpeter, E.E. \& Stahler, S.W., {\it Astrophys.J.}, {\bf 271} (1983), 632.

\item{}Press, E.H. \& Schechter, P., {\it Astrophys.J.}, {\bf 187} (1974), 425.

\item{}Peebles, P.J.E.,{\it Astrophys.J.}, {\bf 155} (1969), 393.

\item{}Rees, M.J. \& Ostriker, J.P. {\it Astrophys.J.}, {\bf 179} (1977), 541.

\item{}Spitzer, L. Jr. {\it The Physics of Fully ionized Gases(New York:Interscience)} (1956), p.88

\item{}Shapiro, P.R., \& Kang, H., {\it Astrophys.J.}, {\bf 318} (1987), 32.

\item{}Susa, H., Sasaki, M. \& Tanaka, T., {\it Prog.Theor.Phys.}, {\bf 92} (1994), 961. 

\item{}White, S.D.M., {\it Astrophys.J.}, {\bf 286} (1984), 38.

\item{}Yamashita, K., {\it Prog.Theor.Phys}, {\bf 89} (1993), 355.

\end{description}
\newpage
\begin{figure}[htb]
   \centerline{\epsfysize 8cm \epsfxsize 14cm \epsfbox{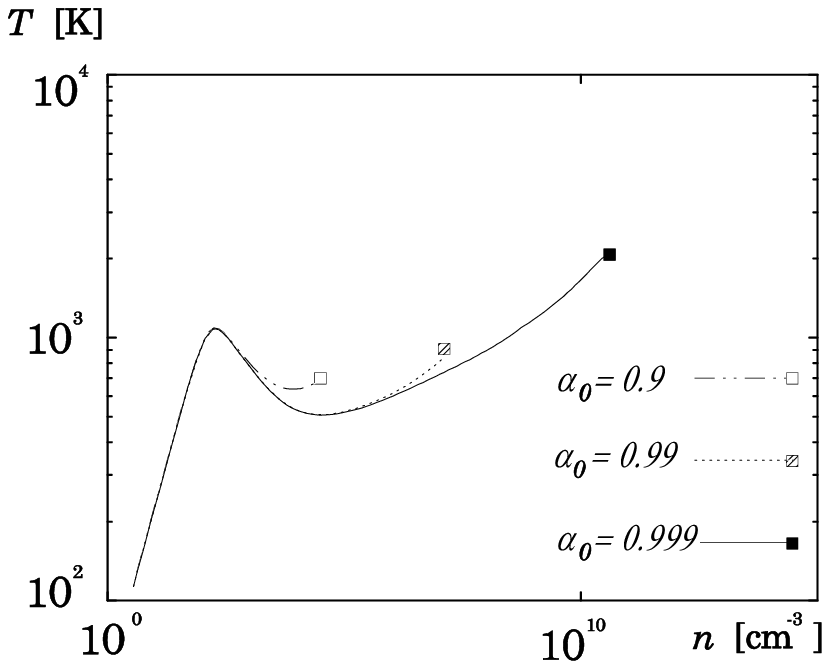}\hspace{2cm}}
   \vspace{0.5cm}
   \caption{
The thermal evolution of the initially oblate clouds on the 
$n-T$ plane. Three lines are corresponding to the initial oblateness,
$\alpha_0 = 0.99, 0.999 \; {\rm and } \; 0.9999$.
   }
   \protect\label{fig1}
\end{figure}
\begin{figure}[htb]
   \centerline{\epsfysize 8cm \epsfxsize 14cm \epsfbox{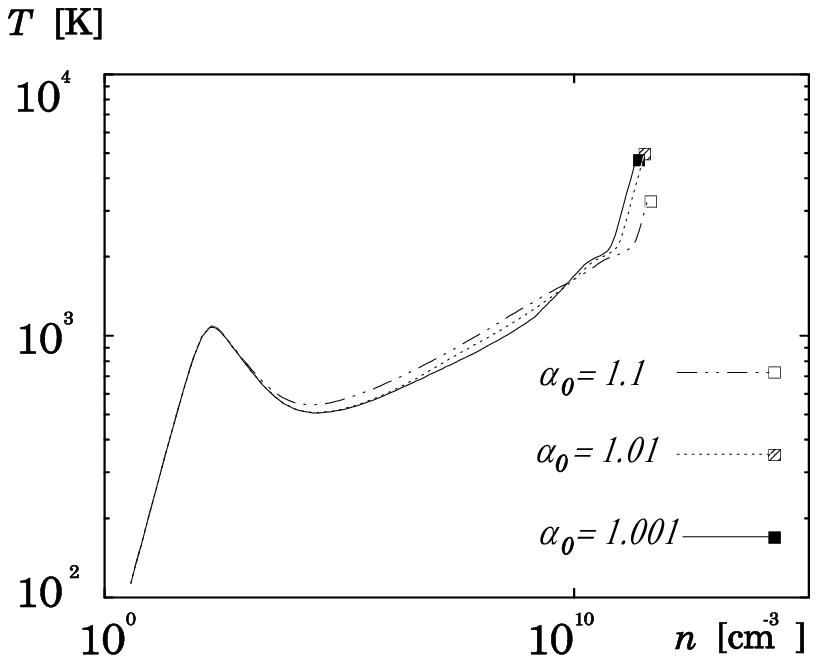}\hspace{2cm}}
   \vspace{0.5cm}
   \caption{
The thermal evolution of the initially prolate clouds on the 
$n-T$ plane. Four lines are corresponding to the initial prolateness,
$\alpha_0 = 1.01, 1.001, 1.0001\; {\rm and } \; 1.00001$.
   }
   \protect\label{fig2}
\end{figure}
\begin{figure}[htb]
   \centerline{\epsfysize 8cm \epsfxsize 14cm \epsfbox{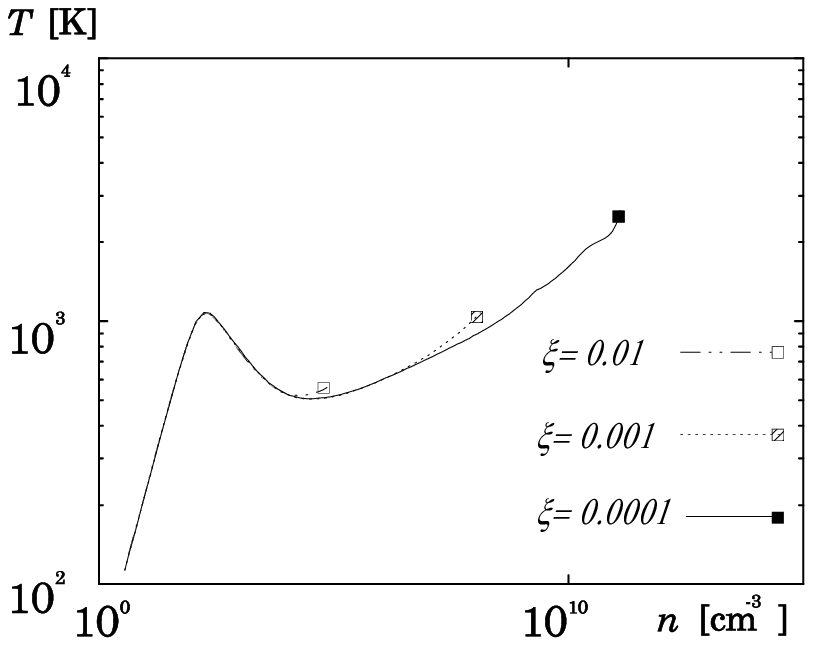}\hspace{2cm}}
   \vspace{0.5cm}
   \caption{
The thermal evolution of the initially spherical clouds with some 
angular momentum on the $n-T$ plane. 
Three lines are corresponding to the magnitude of the initial 
angular momentum, $\xi = 0.01,0.001\; {\rm and } \; 0.0001$.
   }
   \protect\label{fig4}
\end{figure}
\begin{figure}[htb]
   \centerline{\epsfysize 8cm \epsfxsize 14cm \epsfbox{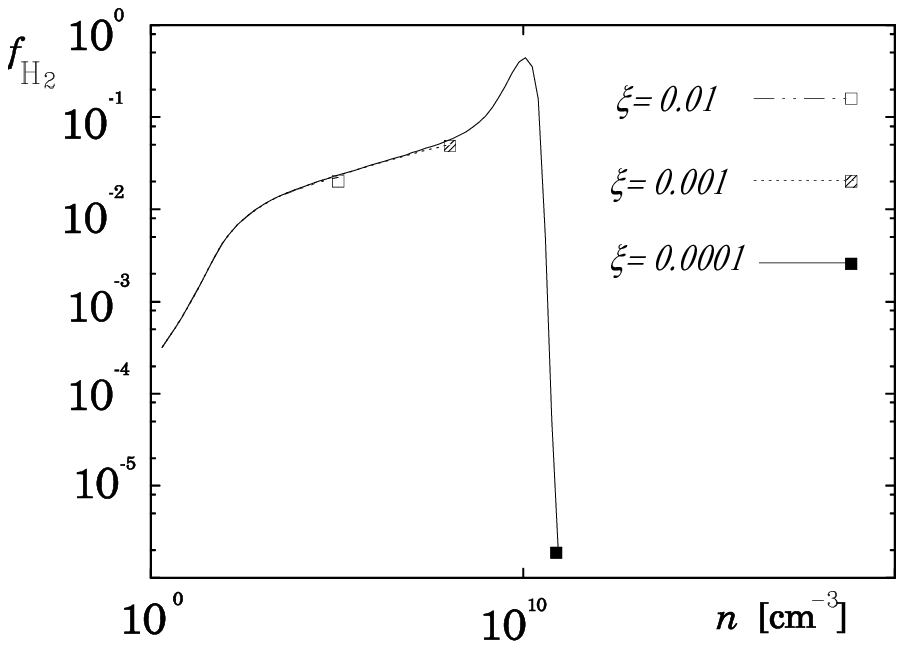}\hspace{2cm}}
   \vspace{0.5cm}
   \caption{
The $H_2$fraction of the initially spherical clouds with some 
angular momentum on the $n-T$ plane. 
Three lines are corresponding to the magnitude of the initial 
angular momentum, $\xi = 0.01,0.001\; {\rm and } \;0.0001$.
   }
   \protect\label{fig3}
\end{figure}
\begin{figure}[htb]
   \centerline{\epsfysize 8cm \epsfxsize 14cm \epsfbox{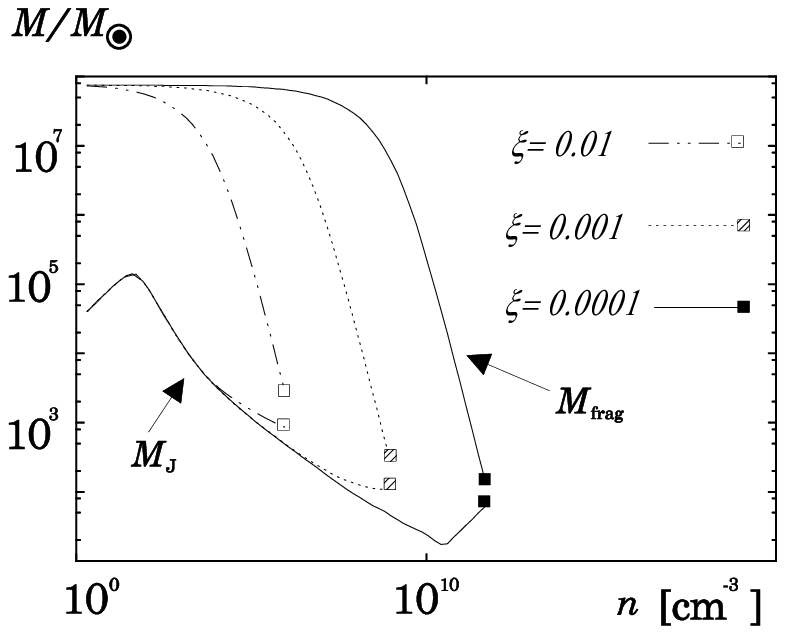}\hspace{2cm}}
   \vspace{0.5cm}
   \caption{
The evolution of the Jeans mass and the mass of the fragmentation
for the initially spherical rotator, in the $n-M$ plane. 
The mass of the fragments are estimated as 
$M_{frag}= 2\rho\lambda_{mgr}^2 $ (see text) . 
}
   \protect\label{fig5}
\end{figure}
\begin{figure}[htb]
   \centerline{\epsfysize 8cm \epsfxsize 14cm \epsfbox{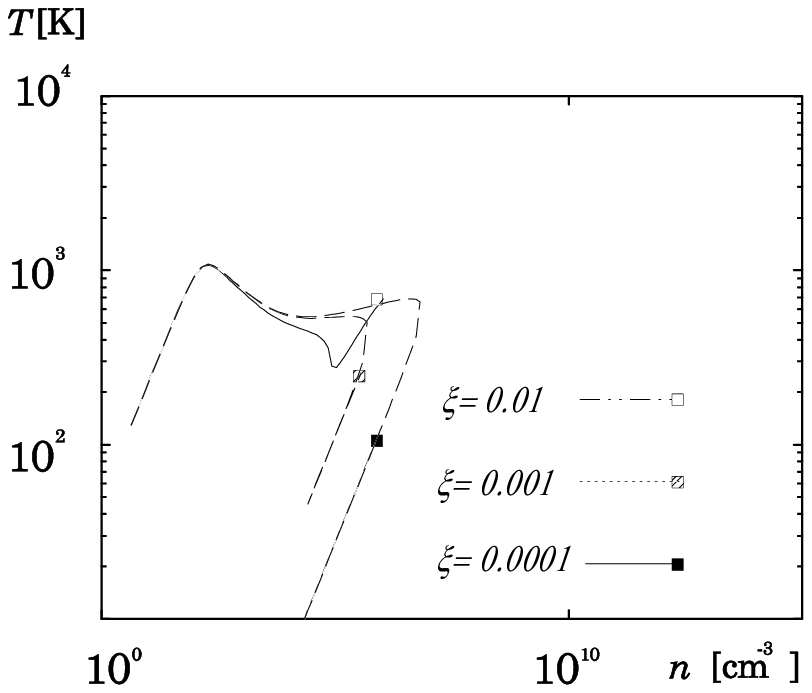}\hspace{2cm}}
   \vspace{0.5cm}
   \caption{
The thermal evolution of the initially prolate clouds with some 
angular momentum on the $n-T$ plane. 
The initial flatness, $\alpha_0$  is $1.1$, and   
three lines are corresponding to the magnitude of the initial 
angular momentum, $\xi = 0.01,0.001\; {\rm and } \; 0.0001$. 
}
   \protect\label{fig6}
\end{figure}
\begin{figure}[htb]
   \centerline{\epsfysize 18cm \epsfxsize 13cm \epsfbox{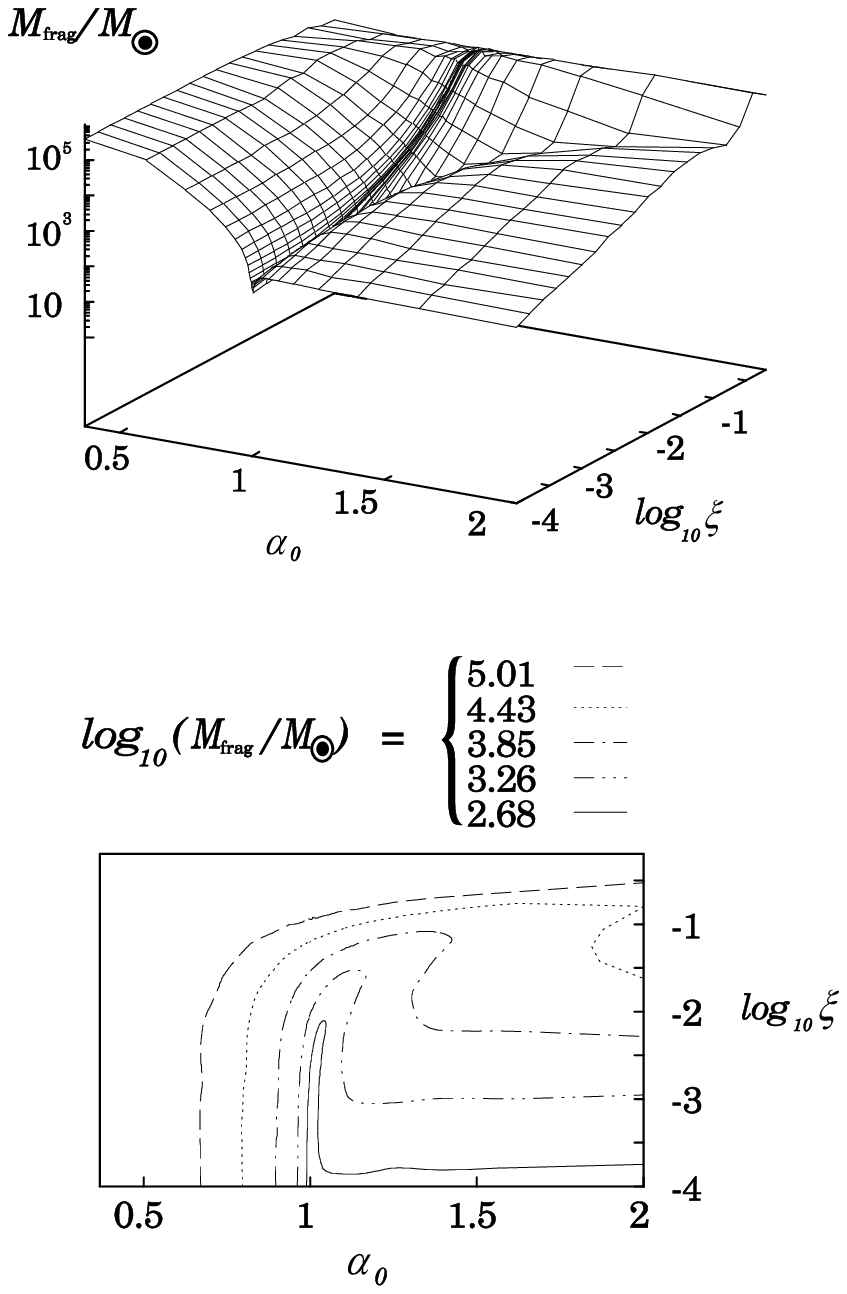}\hspace{2cm}}
   \vspace{0.5cm}
   \caption{The upper figure shows
   $M_{frag}$ for various values of $\alpha_0, \xi$.
   The lower figure shows 
   the contour plot of $M_{frag}$ on the $\alpha_0 - \xi$ plane 
   in the upper panel.
   }
   \protect\label{fig7}
\end{figure}

\end{document}